\documentclass{sf2a-conf2016}
\usepackage{graphicx}
\usepackage{hyperref}
\usepackage[]{natbib}  
\usepackage{epstopdf}
\usepackage{todonotes}

\def\BibTeX{{\rm B\kern-.05em{\sc i\kern-.025em b}\kern-.08em
    T\kern-.1667em\lower.7ex\hbox{E}\kern-.125emX}}
\bibpunct{(}{)}{;}{a}{}{,}  

\newcommand{\teff}{{\mathrm{Teff}}}
\newcommand{\logg}{{\mathrm{\log g}}}

\begin{document}

\TitreGlobal{SF2A 2016}


\title{HR8844: a new hot Am star ?}

\runningtitle{HR8844}

\author{R.Monier$^{1,}$}\address{LESIA, UMR 8109, Observatoire de Paris Meudon, Place J.Janssen, Meudon, France}\address{Lagrange, UMR 7293, Universite de Nice Sophia, Nice, France}
\author{M.Gebran}\address{Department of Physics and Astronomy, Notre Dame University - Louaize, PO Box 72, Zouk Mikael, Lebanon}
\author{F.Royer}\address{GEPI, UMR 8111, Observatoire de Paris Meudon, Place J.Janssen, Meudon, France}


\setcounter{page}{237}


\maketitle


\begin{abstract}
Using one archival high dispersion high quality spectrum of HR8844 (A0V) obtained with the echelle spectrograph SOPHIE at Observatoire de Haute Provence, we show that this star is not a superficially normal A0V star as hitherto thought. The model atmosphere and spectrum synthesis modeling of the spectrum of HR8844 reveals large departures of its abundances from the solar composition. We report here on our first determinations of the elemental abundances of 41 elements in the atmosphere of HR8844. Most of the light elements are underabundant whereas the very heavy elements are overabundant in HR8844. This interesting new chemically peculiar star could be a hybrid object between the HgMn stars and the Am stars.
\end{abstract}

\begin{keywords}
stars: individual, stars: Chemically Peculiar
\end{keywords}


\section{Introduction}
 HR8844 currently assigned an A0V spectral type, is one of the 47 northern slowly rotating early-A stars studied by \cite{Royer14}. 
 This star has been little studied: only 32 references can be found in ADS although it is fairly bright (V=5.89).
 The low projected rotational velocity of HR8844 can either be due to i) a very low inclination angle ($i \simeq 0$) or ii) a very low equatorial velocity $v_{e}$ . In this second case, the star could develop large over and underabundances and be a new Chemically Peculiar (CP) star.
 We have recently synthesized several lines of 41 elements present in the SOPHIE spectrum of HR8844  using model atmospheres and spectrum synthesis including hyperfine structure of various isotopes when necessary. These synthetic spectra were iteratively adjusted to the archival high resolution high signal-to-noise spectrum of HR8844 in order to derive the abundances of these elements. This abundance analysis yields underabundances of the light elements (He, C, N and O) and overabundances of  the iron-peak elements and of the very heavy elements (VHE whose atomic number Z is greater than 30). This  definitely shows that HR884 should be reclassified as a new CP star. We present here preliminary determinations of the elemental abundances in HR8844.
  
\section{Observations and reduction}

HD 30085 has been observed at the Observatoire de Haute Provence using the High Resolution (R =75000) mode of SOPHIE in August 2009.
One 10 minutes exposures was secured with a $\frac{S}{N}$ ratio of about 269. 
We did not observe HR8844 ourselves but fetched the spectrum from the SOPHIE archive.

\section{Model atmospheres and spectrum synthesis }

The effective temperature and surface gravity of HR8844 were first evaluated using Napiwotzky et al's (1993) UVBYBETA calibration of Stromgren's photometry.
The found effective temperature $\teff$ is 9750 $\pm$ 200 K and the surface gravity $\logg$ is 3.80 $\pm$ 0.25 dex. 

A plane parallel model atmosphere assuming radiative equilibrium, hydrostatic equilibrium and local thermodynamical equilibrium has been first computed using the ATLAS9 code \citep{Kurucz92}, specifically the linux version using the new ODFs maintained by F. Castelli on her website\footnote{http://www.oact.inaf.it/castelli/}. The linelist was built starting from Kurucz's (1992) gfhyperall.dat file  \footnote{http://kurucz.harvard.edu/linelists/} which includes hyperfine splitting levels.
This first linelist was then upgraded using the NIST Atomic Spectra Database 
\footnote{http://physics.nist.gov/cgi-bin/AtData/linesform} and the VALD database operated at Uppsala University \citep{kupka2000}\footnote{http://vald.astro.uu.se/~vald/php/vald.php}.
A grid of synthetic spectra was then computed with SYNSPEC48 \citep{Hubeny92} to model the lines. The synthetic spectrum was then convolved with a gaussian instrumental profile and a parabolic rotation profile using the routine ROTIN3 provided along with SYNSPEC48.
We adopted a projected apparent rotational velocity $v_{e} \sin i =  27$ km.s$^{-1}$ and a radial velocity $v_{rad} = 4.50 $ km.s$^{-1}$ from \cite{Royer14}.


\vskip 1cm

\begin{figure}[ht!]
 \centering
 \includegraphics[width=0.8\textwidth]{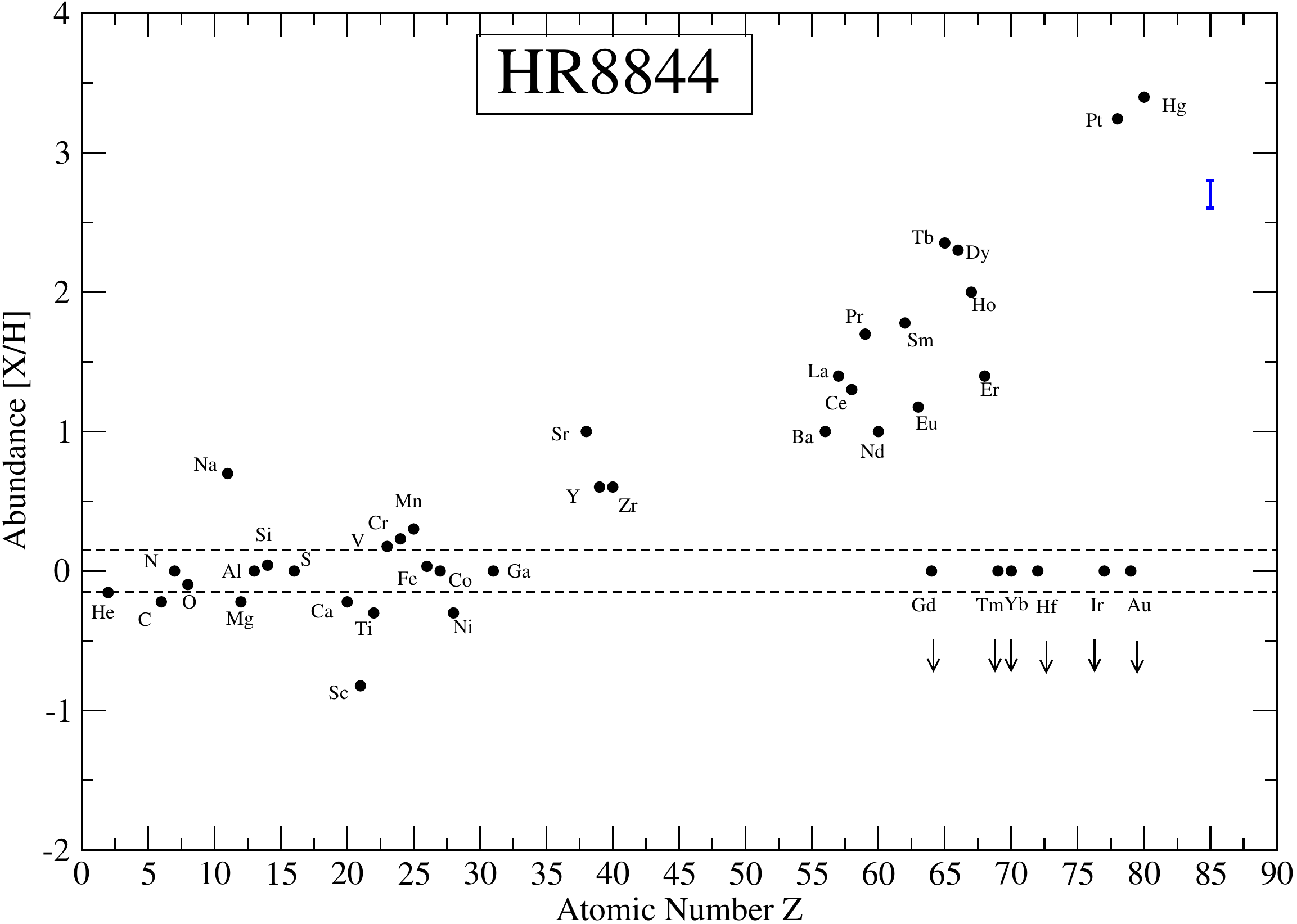}     
  \caption{The derived elemental abundances for HR8844}
  \label{fig1}
\end{figure}

\section{The derived abundance pattern for HR8844}

The derived abundances for the 41 elements studied are displayed in Fig.~\ref{fig1}. For a given element, we display  actually the difference of the abundance in HR8844 relative to the solar value. A null value therefore means a solar abundance, a negative value an underabundance and a positive value an overabundance for that element in HR8844. We have depicted 2 horizontal lines at $\pm$ 0.15 dex of the null values to display a representative error bar. Any abundance situated above or below these lines represent real over or underabundances.  We find that HR8844 displays underabundances in the light elements He, C, O, Mg, Ca, Sc, Ti and Ni. It has solar abundances for N, Al, Si, S and Fe and only mild overabundances for V, Cr, Mn. It has large overabundances in several very heavy elements:Sr, Y, Zr, Ba, La, Ce, Pr, Nd, Sm, Eu, Tb, Dy, Er, Ho, Pt and Hg. The heaviest elements Pt and Hg seem to be the most overabundant however their abundance determinations should be taken with caution as they rely on one line only for each element. The abundance pattern of HR8844 somehow resembles that of the hot Am stars, Sirius A and HD72660 which have effective temperatures and surface gravities very close to that of HR8844. However much work remains to be done to establish the differences and similarities of surface composition in these three stars.

\section{Conclusions}

The derived abundance pattern of HR8844 strongly departs from the solar composition which definitely shows that HR8844 is not a superficially normal early A star but is actually another new CP star. We have already reported on the discovery of 5 new CP stars of the HgMn type in \cite{Monier15} and \cite{Monier2016}. HR8844 has overabundances of both the rare earths and possibly of the VHE Hg and Pt and therefore resembles both a very hot Am star and a very cool HgMn star. It could be a hybrid object intermediate between these two classes of objects.
We are currently planning more observations of HR8844 with SOPHIE in order  to complement the abundances derived here and search for line variability. This will help us adddress the relationship of HR8844 to the 2 other hot Am stars, Sirius A and HD72660 and constrain the nature of this interesting new CP star.

\begin{acknowledgements}
The authors acknowledge use of the SOPHIE archive (\url{http://atlas.obs-hp.fr/sophie/}) at Observatoire de Haute Provence. They have used the NIST Atomic Spectra Database and the VALD database operated at Uppsala University (Kupka et al., 2000) to upgrade atomic data.
\end{acknowledgements}



%
\end{document}